\documentclass[10pt,aps,prd,preprint,showpacs,groupedaddress,nofootinbib]{revtex4}
\usepackage{amsmath,amssymb,amsbsy,color}

\newcommand{\nn}{\nonumber}
\def\dfrac#1#2{\displaystyle\frac{#1}{#2}}

\newcommand{\p}{\partial}
\newcommand{\pslash}{p\kern-1ex /}
\newcommand{\lslash}{l\kern-1ex /}
\newcommand{\sslash}{s\kern-1ex /}
\newcommand{\Dslash}{{\cal D}\kern-1.5ex /}

\newcommand{\tr}{{\rm tr}}

\newcommand{\beqa}{\begin{eqnarray}}
\newcommand{\eeqa}{\end{eqnarray}}

\newcommand{\be}{\begin{equation}}
\newcommand{\ee}{\end{equation}}
\newcommand{\bea}{\begin{eqnarray}}
\newcommand{\eea}{\end{eqnarray}}
\newcommand{\ba}{\begin{array}}
\newcommand{\ea}{\end{array}}

\newcommand{\pref}[1]{(\ref{#1})}
\def\chpt{ChPT}

\begin{document}

\preprint{UTHEP-510}
\preprint{UTCCS-P-16}
\title{
Pseudo scalar meson masses in Wilson Chiral Perturbation Theory for 2+1  
flavors
}
\author{S.~Aoki}
\affiliation{Graduate School of Pure and Applied Sciences, University  
of Tsukuba, Tsukuba, Ibaraki  305-8571, Japan}
\affiliation{
Riken BNL Research Center, Brookhaven National Laboratory, Upton, New York 11973,
USA}
\author{O.~B\"ar}
\affiliation{Graduate School of Pure and Applied Sciences, University  
of Tsukuba, Tsukuba, Ibaraki  305-8571, Japan}

\author{T.~Ishikawa}

\affiliation{
Center for Computational Physics, University of Tsukuba, Tsukuba  
305-8577, Japan
}
\author{S.~Takeda}
\affiliation{Graduate School of Pure and Applied Sciences, University  
of Tsukuba, Tsukuba, Ibaraki  305-8571, Japan}

\date{\today}
%
\begin{abstract}
%
We consider 2+1 flavor Wilson Chiral Perturbation Theory including the  
lattice spacing contributions of O($a^{2}$). We adopt a power counting  
appropriate for the unquenched lattice simulations carried out by the  
CP-PACS/JLQCD collaboration and compute the pseudo scalar meson masses  
to one loop.
These expression are required to perform the chiral extrapolation of  
the CP-PACS/JLQCD lattice data.
\end{abstract}

\maketitle

%
\section{Introduction}
%

The limitations of the quenched approximation in numerical lattice QCD  
simulations is by now well established. For example, the light hadron  
mass spectrum calculated by the CP-PACS collaboration  
\cite{Aoki:2002fd} deviates from the experimentally measured values by  
about 10 percent. Even though the quenching error is different for  
different observables, one must assume the quenching error to be of  
the same order for other quantities as well.
Once the effect of dynamical up and down quarks is included, the  
quenching error  is significantly reduced and the discrepancy between  
the numerically calculated and the experimentally measured values is  
much smaller  compared with the quenched results \cite{AliKhan:2001tx}.  
Still, ignoring the effect of a dynamical strange quark in these  
unquenched 2 flavor simulations  leads to an uncertainty, which is  
expected to be non negligible. Only simulations with a dynamical  
strange quark will provide numerical results, which can be compared with  
experiment with confidence.

In order to eliminate the remaining source of quenching error the  
CP-PACS and JLQCD collaborations have been carrying out  
unquenched 2+1 flavor simulations. A RG-improved gauge action and an  
O($a$) improved Wilson quark action have been adopted. The size of the  
lattice is modest ($L\,\simeq 2$ fm) and simulations at three different  
lattice spacings ($a\simeq 0.7$ fm, $1.0$ fm, $1.22$ fm) are planned so  
that the continuum limit can be taken. Five different masses for the  
degenerate up and down type quarks are simulated leading to pseudo  
scalar meson masses in the range $m_{PS}/m_{V}\simeq 0.62 - 0.78$. The  
physical strange quark mass lies between the two simulated strange  
quark masses and can therefore be reached by an interpolation.  More  
details and the status of these simulations have been recently  
summarized in Ref.\ \cite{Ishikawa:2004xq}.

The masses for the up and down quarks are rather heavy and an  
extrapolation to their physical values is required. The functional  
forms for the extrapolation are usually given by Chiral Perturbation  
Theory (\chpt). In its standard form \cite{Gasser:1983yg,Gasser:1984gg}  
the expressions derived in \chpt\ can be used after the continuum limit  
of the lattice data has been taken. However, for various reasons it is  
advantageous to perform the chiral extrapolation before the continuum  
limit.
In this case \chpt\ needs to  be formulated for lattice QCD at   
non-zero lattice spacing $a$. The main idea how this can be done was  
proposed in Ref.\ \cite{Sharpe:1998xm,Lee:1999zx}. Since then many  
observables have been calculated at one loop order (for an overview  
see Ref.\ \cite{Bar:2004xp}). For lattice theories with Wilson  
fermions, however, all results were derived for unquenched 2 flavor  
Wilson \chpt\ (W\chpt) and the chiral expressions are therefore not  
applicable for the CP-PACS/JLQCD simulations.

This is the first paper in a series where we provide the one loop  
expressions of 2+1 flavor W\chpt\ for a variety of observables, which  
will be measured by the CP-PACS/JLQCD collaboration. Here we present  
the results for the simplest observables, the pseudo scalar meson  
masses. The second paper  is devoted to the vector meson masses and the  
third to the pseudo scalar decay constants and axial vector Ward  
identity quark mass \cite{Aoki2,Aoki3}. In our calculations we include the  
lattice spacing contributions through O($a^{2}$) and adopt various power  
countings. Even though we have primarily the CP-PACS/JLQCD simulations in mind, our expressions are equally useful for other unquenched  2+1 flavor simulations employing Wilson fermions.

There is no fundamental difficulty in applying the framework of WChPT to 2+1 flavors, the main difference to the 2 flavor results is just the increased complexity of the final results. Since we follow the standard strategy of WChPT we will be brief in presenting our results. In section  \ref{sect:CL} we explain the power counting which we assume and present the chiral Lagrangian up to next-to-leading order.
The calculation of the pseudo scalar masses from this Lagrangian is straightforward and we summarize our final results in section \ref{sect:results}. Many technical details and some intermediate results are collected in appendices \ref{sec:appA} and  \ref{sec:appB}.

%
\section{The chiral effective Lagrangian}
\label{sect:CL}
%
\subsection{The order counting}
In continuum chiral perturbation theory (ChPT),
$M= m$ or $p^2$ is the expansion parameter,
where $m$ is the quark mass and $p$ is the momentum of the pseudo scalar meson.
Since chiral symmetry is explicitly broken in lattice QCD with
Wilson-type quarks, corrections due to the non-zero lattice spacing $a$
are non-negligible. 
The construction of the chiral effective theory for Lattice QCD with Wilson fermions, so-called Wilson ChPT (WChPT),  has become standard by now. From a conceptional point of view there is nothing new in applying the familiar techniques \cite{Sharpe:1998xm,Rupak:2002sm,Bar:2003mh,Aoki:2003yv} to 2+1 flavor lattice QCD. The only non-trivial choice one has to make is the order counting one adopts in the chiral expansion, which we are going to explain in this section.\footnote{For notational simplicity we assume $N$ degenerate quark masses in the following discussion.}

The leading order (LO) chiral Lagrangian of the WChPT
is constructed from O$(M)$ terms and the O$(a)$ term.
Since the O$(a)$ term has the same chiral structure as the mass term,
the LO Lagrangian of the WChPT assumes the same form as the one of continuum 
ChPT, provided one performs the replacement $m \rightarrow  \tilde m = m +c_{1}a$ with $c_{1}$ being  a combination of two low-energy constants \cite{Sharpe:1998xm,Rupak:2002sm}.
Based on this LO Lagrangian, however, 
the pion becomes a tachyon for $ \tilde m = m +c_{1}a < 0$ since
the pion mass squared $m_\pi$ is given by $m_\pi^2 = 2 B ( m + c_1 a)$
at tree level. This is in contrast to continuum ChPT where chiral symmetry dictates that
the pion mass is given by $m_\pi^2 = B \vert m \vert$.
This already indicates that the first non-trivial modification to continuum ChPT starts at O$(a^2)$. 
Indeed, including the O$(a^2)$ term in the LO chiral Lagrangian removes the unphysical tachyon from the theory, as has been shown in Ref. \cite{Aoki:2003yv} for the 2 flavor theory.
We therefore conclude that, for the consistency of the theory, the O$(a^2)$ term should be included in the LO  Lagrangian in the WChPT.

Although the O$(a)$ term is superficially larger than the O$(a^2)$ term,
the former term is irrelevant since it is absorbed in the definition of the shifted mass  $\tilde m$.
The O$(a^2)$ correction, on the other hand, becomes important in the regime where
$\tilde m/\Lambda_{\rm QCD}\simeq
\Lambda_{\rm QCD}^2 a^2 $, even though the 
$\Lambda_{\rm QCD}^2 a^2$ correction is much smaller than 1
in general.

Suppose we consider the O$(\tilde M) =O(\tilde m, p^{2})$ and the O$(a^2)$ term as LO terms.
Then the terms of O$(\tilde M^2,\tilde M a^2,a^{4})$ can be regarded as next to leading order (NLO), since the loop expansion in WChPT  
increases in units of $\tilde M$.
We remark that the O$(a^{4})$ term is not as relevant as the O$(\tilde M^2,\tilde M a^2)$ terms for our final results for the pseudo scalar masses. 
We will need the tree level contribution of the O$(a^4)$ term
only to satisfy $m_{\rm PS,NLO}^2 = 0$ at $m_{\rm PS,LO}^2 = 0$, where
$m_{\rm PS,NLO}$ is the pseudo scalar meson mass at  NLO.
This means that there is essentially no unknown low energy constant associated with the  O$(a^4)$ correction.\footnote{This may seem odd within WChPT, but it is a simple consequence of the fact that the pseudo scalar meson becomes massless in lattice QCD at the critical quark mass. In other words, the O$(a^4)$ term merely results in an additional shift in the critical quark mass. An implicit assumption we make here is the presence of a phase where flavor and parity is spontaneously broken \cite{Aoki:1983qi}. }

The proper order counting of the O$( \tilde M a)$ term is more subtle than for the previously discussed terms.\footnote{The O$(a^{3})$ term is unproblematic since the arguments we gave for the O$(a^{4})$ contribution also apply for the O$(a^{3})$ term.}
Depending on the size of the O$(\tilde M a)$ contributions we 
may include it at leading order where it subsequently enters the chiral logs, or we treat it as a NLO term and include it at tree level only. 
The size of this term is indeed expected to vary significantly, depending on details of the action of the underlying lattice theory. The O$(\tilde M a)$ term contains one power of $a$ and stems from the Pauli term in the Symanzik's effective action \cite{Symanzik:1983dc,Symanzik:1983gh,Sheikholeslami:1985ij}, which is used in an intermediate step to match the lattice to the chiral effective theory. Consequently, the O$(\tilde M a)$ contribution in the chiral Lagrangian is directly proportional to the coefficient of the Pauli term \cite{Rupak:2002sm}, and the size of this coefficient is much smaller for improved theories with a clover term in the action than for standard Wilson fermions. For fully non-perturbatively improved theories the coefficient is equal to zero and no O$(\tilde M a)$ term appears in the chiral Lagrangian.\footnote{Strictly speaking this only holds if the chiral Lagrangian is parameterized in terms of renormalized quark masses which absorb some of the O($a$) cut-off effects. Particular terms linear in $a$ will appear if other choices are used (see section \ref{Oaimprovedtheories}).} 

Since we have no {\em a priori} knowledge about the size of the O$(\tilde M a)$ term we will be as general as possible and present our results for three different order countings of this term. We first keep the O$(\tilde M a)$ term at LO where it gives a contribution to the chiral log corrections. If the O$(\tilde M a)$ term is assumed to be much smaller than the other LO terms we can easily expand our results. Performing this expansion is equivalent to doing the calculation with keeping the O$(\tilde M a)$ term at 
NLO. Finally we can set this term to zero in order to obtain the results for non-perturbatively improved theories. 
%
\subsection{Leading order Lagrangian}
%
According to the discussion in the previous section, we use the following LO Lagrangian, which consists of terms of
O$(\tilde M, a^2, \tilde Ma )$:
\beqa
L_{\rm LO} &=&
\frac{f^2}{4}\langle \partial_\mu \Sigma  
\partial_\mu\Sigma^\dagger\rangle
-\frac{f^2 B}{2}\langle M_q\Sigma + \Sigma^\dagger M_q\rangle \nn\\
&+&\frac{f^2}{16}\left[ c_2 \langle \Sigma +\Sigma^\dagger\rangle^2
+ \tilde c_2 \langle (\Sigma +\Sigma^\dagger)^2 \rangle
+ c_4 \langle \Sigma -\Sigma^\dagger\rangle^2
 \right] \nn\\
&+&\frac{f^2}{4}\left[ c_0 \langle \Sigma +\Sigma^\dagger-2\rangle
\langle \partial_\mu \Sigma \partial_\mu\Sigma^\dagger\rangle + \tilde  
c_0
\langle (\Sigma +\Sigma^\dagger-2) \partial_\mu
\Sigma\partial_\mu\Sigma^\dagger \rangle \right] \nn\\
&+&\frac{f^2 B}{8}\left[ 2 c_3 \langle \Sigma +\Sigma^\dagger\rangle
\langle M_q\Sigma + \Sigma^\dagger M_q\rangle
+ \tilde c_3 \langle (\Sigma +\Sigma^\dagger )( M_q\Sigma +  
\Sigma^\dagger M_q)
\rangle\right]\nn\\
&+&\frac{f^2 B}{8}\left[2 c_5 \langle \Sigma -\Sigma^\dagger\rangle
\langle M_q\Sigma - \Sigma^\dagger M_q\rangle
+ \tilde c_5 \langle (\Sigma -\Sigma^\dagger )
( M_q\Sigma - \Sigma^\dagger M_q)\rangle\right],
\label{eq:LO}
\eeqa
where $\langle X \rangle = \tr\ X$, $f$ is the pseudo scalar meson decay constant,
\beqa
\Sigma &=& \exp\left[ i\frac{1}{f}\sum_a \pi_a T^a\right],
\eeqa
is an element of SU(3) with $\pi_a$ being the pseudo scalar meson fields. The SU(3) generators $T^a$ are normalized according to $\tr\ T^a T^b =\frac{1}{2}\delta_{ab}$.
The first and the second term in the first line are the standard O$(p^2)$ and  
O$(\tilde m)$
terms \cite{Gasser:1984gg}, respectively.  The second line comprises the O$(a^2)$ terms \cite{Bar:2003mh,Aoki:2003yv}. The last three lines contain the O$( p^2 a)$ and  O$( \tilde m a)$ contributions \cite{Rupak:2002sm}.

For notational simplicity only we use a different notation for the low energy constants associated with the non-zero lattice spacing (the $c$ and $\tilde c$'s) compared to the notation in Refs.\  \cite{Rupak:2002sm,Bar:2003mh,Aoki:2003yv}. In particular, we have chosen to absorb the explicit powers of the lattice spacing $a$ into the coefficients $c, \tilde c$. Consequently, as a function of $ a$ these coefficients scale according to 
\bea
c_{0}, \tilde c_{0}, c_{3}, \tilde c_{3}, c_{5}, \tilde c_{5} & = & {\rm O} (a),\qquad
c_{2}, \tilde c_{2}, c_{4} \, =  \, {\rm O} (a^{2}).
\eea
The quark mass matrix is given by
\beqa
M_q &=& \left(
\begin{array}{ccc}
m & 0 & 0 \\
0 & m & 0 \\
0 & 0 & m_s\\
\end{array}
\right) = M_0 {\bf 1} + M_8 T^8, \quad
M_0 =\frac{2m + m_s}{3},\quad M_8 =\frac{2(m-m_s)}{\sqrt{3}} ,
\eeqa
where isospin symmetry ($m_u=m_d=m$) is assumed.
Note that  O$(a)$ contribution is already absorbed in the definition of $M_q$, so there is no O($a$) term in the chiral Lagrangian.
In the $a\rightarrow 0$ limit, the pseudo scalar meson masses are related to $m$ and  
$m_s$ according to 
\beqa
m_a^2 &=&\left\{
\begin{array}{ll}
m_\pi^2 = 2 B m, & a=1,2,3, \\
m_K^2 = B (m+m_s), & a=4,5,6,7, \\
m_\eta^2 = \dfrac{2B}{3}(m+2m_s), & a=8,
\end{array}\right.
\eeqa
which, of course, agree with the continuum \chpt\ result. We also define the average mass
\beqa
m_{\rm av}^2 & =& \frac{1}{N^2-1}\sum_a m_a^2 = 2B\frac{ 2m + m_s}{3},
\eeqa
which will be a useful short hand notation in the following.

Note that, except for a $\pi_{a}$ independent constant,
the term proportional to  $\tilde c_5$ is identical to the
term with $\tilde c_3$. Therefore we can set $\tilde c_5=0$ without loss of generality.
%
\subsection{Shifted quark mass at leading order}
%
By expanding the LO Lagrangian to second order in $\pi_a$, we obtain
\beqa
L^{(2)}
&=&\frac{1}{2}\sum_a\left[ (\partial_\mu\pi_a)^2
+\tilde m_a^2 \pi_a^2 \right].
\label{eq:LO_2pt}
\eeqa
The pseudo scalar meson masses at  LO are therefore given by
\beqa
\tilde m_a^2 &=& m_a^2(1-N c_3-\tilde c_3)-m_{\rm av}^2 N c_3
-Nc_2-\tilde c_2.
\eeqa
In the following we keep the number of flavors $N$ arbitrary, but put $N=3$ in the final expressions.

We now define {\em shifted quark masses}, which satisfy
\beqa
\tilde m_a^2 &=& \left\{
\begin{array}{ll}
\tilde m_\pi^2 = 2 B \tilde m, & a=1,2,3, \\
\tilde m_K^2 = B (\tilde m+\tilde m_s), & a=4,5,6,7,\\
\tilde m_\eta^2 = \dfrac{2B}{3}(\tilde m+2\tilde m_s), & a=8.\\
\end{array}
\right. \\
\eeqa
Explicitly they are given by
\beqa
\tilde m &=& m (1-N c_3-\tilde c_3)-\frac{2m+m_s}{3} N c_3
-\frac{Nc_2+\tilde c_2}{2B}, \\
\tilde m_s &=& m_s (1-N c_3-\tilde c_3)-\frac{2m+m_s}{3} N c_3
-\frac{Nc_2+\tilde c_2}{2B} .
\eeqa
From these expressions the LO critical mass for $N=3$ is
defined  by the condition
\beqa
\tilde m_{\rm av}^2 &=& m_{\rm av}^2(1-6c_3-\tilde c_3)
-3c_2-\tilde c_2 = 0,
\eeqa
leading to
\beqa
2 B m_{\rm critical} &=& \frac{3c_2+\tilde c_2}{1-6c_3-\tilde c_3}
\equiv -\delta m_{\rm av}^2.
\eeqa
This definition for the critical quark mass assumes that all three quark masses are extrapolated to the massless point, including the strange quark mass. In numerical spectroscopy calculations, however, a different definition is sometimes employed where the strange quark mass is kept fixed at (approximately) its physical value. For this procedure the condition for the critical quark mass reads
\beqa
\tilde m_\pi^2 &=& m_\pi^2 (1-3c_3-\tilde c_3)-m_{\rm av}^2 3 c_3  
-3c_2-
\tilde c_2 \,=\, 0,
\eeqa
which results in
\beqa
2 B m_{\rm critical}(m_s) &=&
\frac{2Bm_s c_3 +3c_2+\tilde c_2}{1-5c_3-\tilde c_3}.
\eeqa
The difference between these two values is therefore
\beqa
m_{\rm critical} - m_{\rm critical }(m_s)
&=&
\frac{c_3}{1-5c_3-\tilde c_3}\left[
\frac{3c_2+\tilde c_2}{1-6c_3-\tilde c_3}
- 2B m_s
\right].
\eeqa
Indeed, numerical 2+1 flavor simulations \cite{Kaneko:2003xq} suggest a non-vanishing value for this difference with
\beqa
m_{\rm critical} - m_{\rm critical }(m_s) > 0 .
\eeqa

%
\subsection{NLO Lagrangian}
%
The NLO Lagrangian provides the necessary counter terms in order to remove the UV divergences in the 1-loop integrals. The contribution of  O$(\tilde M^2)$ is given by
\beqa\label{NLOMSquared}
L_{{\rm NLO},\tilde M^2}&=&L_4\langle \Sigma_{\mu\mu}\rangle \langle \hat M_q\Sigma
+\Sigma^\dagger \hat M_q\rangle
+ L_5\langle \Sigma_{\mu\mu} (\hat M_q\Sigma +\Sigma^\dagger \hat  
M_q)\rangle
+L_6\langle \hat M_q\Sigma +\Sigma^\dagger \hat M_q\rangle^2 \nn\\ && + \, 
L_7\langle \hat M_q\Sigma -\Sigma^\dagger \hat M_q
\rangle^2
+L_8\langle \hat M_q\Sigma \hat M_q\Sigma + \Sigma^\dagger \hat  
M_q\Sigma^\dagger \hat M_q\rangle,
\eeqa
where we introduced $\Sigma_{\mu\mu} = \partial_\mu \Sigma \partial_\mu\Sigma^\dagger$ and  
$\hat M_q = B \tilde M_q$, where $\tilde M_q$ is the shifted quark mass matrix constructed from $\tilde m$
and $\tilde m_s$ (cf.\ previous section). 
The NLO constants $L_{i}$
are related with standard Gasser-Leutwyler coefficients $L_{i}^{GL}$
\cite{Gasser:1984gg} as
$L_{4,5}=2L_{4,5}^{GL}$ and 
$L_{6,7,8}=-4L_{6,7,8}^{GL}$.

The complete Lagrangian at O$(a^2 \tilde M)$ and O$(a \tilde M^2)$
is cumbersome and has not been computed so far. 
Here we only list the terms that contribute to the meson propagators,
which we need for the calculation of the pseudo scalar masses. These
terms are straightforwardly found by a spurion analysis
with spurion fields proportional to the lattice spacing
\cite{Sharpe:1998xm,Rupak:2002sm,Bar:2003mh}. 
Our result for the O$(a^2 \tilde M)$  Lagrangian reads
\beqa
L_{{\rm NLO}, a^2\tilde M} &=&\langle  
\Sigma_{\mu\mu}\rangle\left(W_0+\frac{W_1}{4N^2}\langle \Sigma  
+\Sigma^\dagger\rangle^2
+\frac{W_2}{2N}\langle \Sigma^2 +(\Sigma^\dagger)^2\rangle\right)\nn\\
&+&\frac{W_3}{4N}\langle \Sigma_{\mu\mu}(\Sigma+\Sigma^\dagger)\rangle  
\langle \Sigma +\Sigma^\dagger\rangle
+\frac{W_4}{2}\langle \Sigma_{\mu\mu}(\Sigma^2+(\Sigma^\dagger)^2)\rangle  
\nn\\
&+& W_5 \langle \hat M_q\Sigma +\Sigma^\dagger \hat M_q\rangle
+W_6 \langle \Sigma +\Sigma^\dagger\rangle^2 \langle \hat M_q\Sigma  
+\Sigma^\dagger \hat M_q\rangle
+W_7 \langle \Sigma^2 +(\Sigma^\dagger)^2\rangle \langle \hat M_q\Sigma  
+\Sigma^\dagger \hat M_q\rangle\nn\\
&+&W_8 \langle \Sigma +\Sigma^\dagger\rangle \langle \hat M_q\Sigma^2  
+(\Sigma^\dagger)^2 \hat M_q\rangle
+ W_9 \langle \hat M_q\Sigma^3 +(\Sigma^\dagger)^3 \hat M_q\rangle
+ W_{10}\langle \Sigma +\Sigma^\dagger\rangle \langle \hat  
M_q\rangle\nn\\
&+&W_{11}\langle (\p_\mu\Sigma)^2 +(\p_\mu\Sigma^\dagger)^2\rangle +
W_{12}\langle (\p_\mu\Sigma)^2(\Sigma^\dagger)^2 +
\Sigma^2(\p_\mu\Sigma^\dagger)^2\rangle .
\eeqa
while the terms of O$(a\tilde M^2)$ are given by
\beqa
L_{{\rm NLO},a\tilde M^2}&=& \langle \Sigma_{\mu\mu}\rangle
\left[2 V_0 \langle \hat M_q\rangle + \frac{V_1}{2N}\langle \Sigma  
+\Sigma^\dagger\rangle
\langle \hat M_q\Sigma +\Sigma^\dagger \hat M_q\rangle
+ V_2 \langle \hat M_q\Sigma^2 +(\Sigma^\dagger)^2 \hat  
M_q\rangle\right]\nn\\
&+& \frac{V_3}{2} \langle \Sigma_{\mu\mu}(\Sigma +\Sigma^\dagger)\rangle  
\langle \hat M_q\Sigma
+\Sigma^\dagger \hat M_q\rangle
+ \frac{V_4}{2N}\langle \Sigma_{\mu\mu}(\hat M_q\Sigma +\Sigma^\dagger \hat  
M_q)\rangle
\langle \Sigma +\Sigma^\dagger \rangle \nn\\
&+& V_5 \langle \Sigma_{\mu\mu}(\hat M_q\Sigma^2 +(\Sigma^\dagger)^2 \hat  
M_q)\rangle
+ 2 V_6 \langle \Sigma_{\mu\mu}\hat M_q\rangle \nn\\
&+& V_{16} \langle (\p_{\mu} \Sigma)^2 \hat M_q + \hat M_q (\p_{\mu} \Sigma^{\dag})^2 \rangle
+ V_{17} \langle (\p_{\mu} \Sigma)^2 \Sigma^{\dag} \hat M_q \Sigma^{\dag}
      + \Sigma \hat M_q \Sigma (\p_{\mu} \Sigma^{\dag})^2 \rangle \nn\\
&+& \langle \Sigma +\Sigma^\dagger \rangle\left[ V_7\langle \hat  
M_q^2\rangle
+ V_8 \langle \hat M_q\Sigma +\Sigma^\dagger \hat M_q\rangle^2
+ V_9 \langle \hat M_q\Sigma \hat M_q\Sigma +\Sigma^\dagger \hat  
M_q\Sigma^\dagger \hat M_q \rangle\right]\nn\\
&+& \langle \hat M_q\Sigma +\Sigma^\dagger \hat M_q\rangle \left[
V_{10}\langle \hat M_q\rangle + V_{11} \langle \hat M_q\Sigma^2  
+(\Sigma^\dagger)^2 \hat M_q\rangle\right]
+V_{12}\langle \hat M_q^2\Sigma +\Sigma^\dagger \hat M_q^2\rangle  \nn  
\\
&+& V_{13}\langle \hat M_q\Sigma \hat M_q\Sigma^2 +(\Sigma^\dagger)^2  
\hat M_q\Sigma^\dagger \hat M_q \rangle
+ V_{14}\langle \Sigma +\Sigma^\dagger \rangle \langle \hat M_q\Sigma  
-\Sigma^\dagger \hat M_q\rangle^2 \nn\\
&+& V_{15}\langle \hat M_q\Sigma^2 -(\Sigma^\dagger)^2 \hat M_q\rangle
\langle \hat M_q\Sigma -\Sigma^\dagger \hat M_q\rangle .
\eeqa
As in the leading order Lagrangian we have absorbed the explicit powers of the lattice spacing in the low-energy constants, and their scaling behaviour is therefore
\bea
W_{i}& =& {\rm O}(a^{2}), \quad i\,=\,0\ldots 12\,,\\
V_{i}& =& {\rm O}(a), \quad i\,=\, 0\ldots 17\,.
\eea
%
\section{Results}\label{sect:results}
%
\subsection{Quark mass dependence of meson masses}
\label{subsec:quarkmassdependenceofmesonmasses}
The calculation of the pseudo scalar masses from the chiral Lagrangian in the previous section is straightforward. We collect some details and intermediate results of our calculation in the appendix. Here we simply quote the final result for the quark mass and lattice spacing dependence
of  the pseudo scalar meson masses. The total contribution from LO tree, LO 1-loop plus NLO tree for $m_{\pi}$ and $m_{K}$ and $m_{\eta}$ are
given  as follows:
\beqa
m_{\pi, \rm total}^2 &=& x+2y +\frac{1}{f^2}\left[
  L_\pi^r \{ A^\pi_\pi x + B^\pi_\pi y + 5 C\}
+L_K^r \{ A^\pi_K x + B^\pi_K y + 4 C\}
+L_\eta^r \{ A^\pi_\eta x + B^\pi_\eta y +  C\}\right.\nn \\
&-&\left.
\{ (C_0+D_0)x + 2 C_0 y + (C_{\rm av}+D_{\rm av})x^2 + D^\pi_{yy}y^2+2D  
xy\}
\right],
\label{eq:final_pi}\\
m_{K, \rm total}^2 &=& x-y +\frac{1}{f^2}\left[
  L_K^r \{ A^K_K x + B^K_K y + 6 C\}
+L_\pi^r \{ A^K_\pi x + B^K_\pi y + 3 C\}
+L_\eta^r \{ A^K_\eta x + B^K_\eta y +  C\}\right.\nn\\
&-&\left.
\{ (C_0+D_0)x - C_0 y + (C_{\rm av}+D_{\rm av})x^2 + D^K_{yy}y^2-D  
xy\}\right],
\label{eq:final_K}\\
m_{\eta, \rm total}^2 &=& x-2y +\frac{1}{f^2}\left[
  L_\eta^r \{ A^\eta_\eta x + B^\eta_\eta y + 3 C\}
+L_\pi^r \{ A^\eta_\pi x + B^\eta_\pi y + 3 C\}
+L_K^r \{ A^\eta_K x + B^\eta_K y +  4 C\}\right.\nn\\
&-&\left.
\{ (C_0+D_0)x - 2 C_0 y + (C_{\rm av}+D_{\rm av})x^2 + D^\eta_{yy}y^2-2  
D xy\}
\right] .
\label{eq:final_eta}
\eeqa
The parameters
\beqa
x&=&\tilde m_{\rm av}^2 = \frac{2B}{3}(2\tilde m+\tilde m_s),
\quad
y= \frac{1}{4\sqrt{3}} 2B \tilde M_8 =\frac{2B}{6}(\tilde m-\tilde m_s)  
,
\eeqa
represent the quark mass dependence.
The chiral log is denoted by
\beqa
L_a^r &=&\frac{\tilde m_a^2}{16\pi^2}\log (\tilde m_a^2),
\eeqa
whose coefficients contain
\beqa
C &=& \frac{1}{6}Z, \nn \\
A^\pi_\pi &=& \frac{1}{3}(2C^\pi_{\pi}-\frac{5}{2}X),\quad
A^\pi_K    =  \frac{1}{3}(2C^\pi_{K}-2X),\quad
A^\pi_\eta =  \frac{1}{3}(2C^\pi_{\eta}-\frac{1}{2}X),\nn\\
B^\pi_\pi &=& \frac{1}{3}(4C^\pi_{\pi}-5Y),\quad
B^\pi_K    =  \frac{1}{3}(C^\pi_{K}-Y),\quad
B^\pi_\eta =  \frac{1}{3}(12c_5-Y),\nn\\
A^K_K  &=& \frac{1}{3}(2C^K_{K}-3X),\quad
A^K_\pi =  \frac{1}{3}(2C^K_{\pi}-\frac{3}{2}X),\quad
A^K_\eta=  \frac{1}{3}(2C^K_{\eta}-\frac{1}{2}X)\nn,\\
B^K_K  &=& \frac{1}{3}(-2C^K_{K}+3Y),\quad
B^K_\pi =  \frac{1}{3}(C^K_{\pi}-\frac{3}{4}Y),\quad
B^K_\eta=  \frac{1}{3}(-3C^K_{\eta}+\frac{5}{4}Y-6c_5)\nn,\\
A^\eta_\eta &=& \frac{1}{3}(2C^\eta_{\eta}-\frac{3}{2}X),\quad
A^\eta_\pi   =  \frac{1}{3}(2C^\eta_{\pi}-\frac{3}{2}X),\quad
A^\eta_K     =  \frac{1}{3}(2C^\eta_{K}-2X),\nn\\
B^\eta_\eta &=& \frac{1}{3}(-4C^\eta_{\eta}+5Y-24c_5),\quad
B^\eta_\pi   =  \frac{1}{3}(-3Y+36c_5),\quad
B^\eta_K     =  \frac{1}{3}(-3C^\eta_{K}+5Y-24c_5) .\nn
\eeqa
These terms are parametrized by
\beqa
X&=&\tilde A\left[1-6c_3-4\tilde c_3 -36 c_3^2 \tilde B\right],\
Y=(1-3c_3-4\tilde c_3)\tilde B,\
Z=\left[9c_2+4\tilde c_2
+\delta m_{\rm av}^2(1-18c_3-4\tilde c_3) \right],
\label{eq:1loop_m_param1}
\\
C^\pi_{\pi}&=&2+18c_0+9\tilde c_0 +15 c_3 \tilde B,\
C^\pi_{K}=1+24c_0+6\tilde c_0 +12 c_3 \tilde B, \
C^\pi_{\eta}=6c_0+\tilde c_0 +3 c_3 \tilde B,\\
C^K_{K}&=&\frac{3}{2}+24c_0+9\tilde c_0 +18 c_3\tilde B,\
C^K_{\pi}=\frac{3}{4}+18c_0+\frac{9}{2}\tilde c_0 +9 c_3\tilde B, \
C^K_{\eta}=\frac{3}{4}+6c_0+\frac{5}{2}\tilde c_0 +3 c_3\tilde B,\\
C^\eta_{\eta}&=&6c_0+3\tilde c_0 +9 c_3\tilde B, \
C^\eta_{\pi}=18c_0+3\tilde c_0 +9 c_3\tilde B, \
C^\eta_{K}=3+24c_0+10\tilde c_0 +12 c_3 \tilde B,
\label{eq:1loop_m_param2}
\eeqa
with
\beqa
\tilde A&=& \frac{1}{1-2N c_3 -\tilde c_3}, \quad
\tilde B= \frac{1}{1-N c_3 -\tilde c_3} , \quad
\delta m_{\rm av}^2 = -\frac{Nc_2+\tilde c_2}{1-2Nc_3-\tilde c_3}.
\nn
\eeqa
The polynomial (non-logarithmic) terms contain
\beqa
D^\pi_{yy}&=& 16\tilde L_5+8(\tilde L_8 +\tilde L_8^\prime) + 48 V_{\Delta}, \quad
D^K_{yy}= 4\tilde L_5+20\tilde L_8 - 16\tilde L_8^\prime+ 48 V_{\Delta},\nn\\
D^\eta_{yy}&=& 16\tilde L_5+24\tilde L_8+24\tilde L_8^\prime +  
96\tilde L_7
+ 48 V_{\Delta}, \quad
D = C_{\rm av} +4\tilde L_{5} + 4(\tilde L_8+\tilde L_8^\prime), \nn \\
C_{\rm av} &=& 4(N\tilde L_6 + N\tilde L_4 +\tilde L_5),\quad
D_{\rm av} = 2(\tilde L_8 + \tilde L_8^\prime) + V_{\rm av},\nn \\
C_0 &=& 4(W_0+W_1+W_2+W_3+W_4)+2W_5 +8N^2 W_6 +4NW_7  
+16NW_8+18W_9\nn\\
& &-\, 8(W_{11}+W_{12})=a^2 W_C, \nn\\
D_0 &=& N[ 16(NW_6+W_7)+4W_8 +2W_{10}]=a^2 W_D , \nn
\eeqa
where
\beqa
\tilde L_4 &=& L_4 + V_0+V_1+V_2+V_3 = L_4 +a L_4^1, \quad
\tilde L_5 = L_5 + V_4+V_5+V_6-V_{16}-V_{17} = L_5 +a L_5^1,\nn\\
\tilde L_6 &=& L_6 + 2NV_8+\frac{1}{4}V_{10}+\frac{5}{2}V_{11} = L_6 +a  
L_6^1,
\quad
\tilde L_7 = L_7 + 2NV_{14}+2V_{15} = L_7 +a L_7^1,\nn\\
\tilde L_8 &=& L_8 + 2N V_9+\frac{1}{2}V_{12}+\frac{5}{2}V_{13}
= L_8 +a L_8^{1a}, \quad
\tilde L_8^\prime = L_8 + 2N V_9+2V_{13}= L_8 +a L_8^{1b},\nn \\
V_{\rm av} &=& N(V_7 + 2 V_9 + 4 N V_8), \quad
V_{\Delta} = \frac{1}{2}(V_7 + 2 V_9) . \nn
\eeqa
Even though the NLO parameters have been used to remove the divergent terms from the loop integrals  we use the same notation for these coefficients.
We finally note that there exists the following constraint among some of the coefficients:
\beqa
\sum_b A^a_b &=&\frac{1}{3}[6+96c_0+32\tilde c_0 +60 c_3 B - 5X].
\eeqa
Consequently, in the limit  $y\rightarrow 0$ we obtain identical results for 
$m_{\pi}$, $m_{K}$ or $m_{\eta}$, as it should be for three degenerate quark masses. 

Obviously the final results for the pseudo scalar masses are fairly lengthy. From a practical point of view the number of independent unknown parameters in these expressions is crucial for their usefulness. 
Unknown parameters are the critical quark mass $m_{\rm critical}$, the constant $2 B$ and
the decay constant $f$. The coefficients of the chiral log terms, given in eq.(\ref{eq:1loop_m_param1})-(\ref{eq:1loop_m_param2}),
are defined through  five independent O$(a)$ parameters, $c_0$, $\tilde  
c_0$,
$c_3$, $\tilde c_3$ and $c_5$, and the coefficient $C$, which is an independent
O$(a^2)$ parameter. In the analytic NLO correction we find the independent combinations $C_{\rm av} + D_{\rm av}$, $D^\pi_{yy}$, $D^K_{yy}$, $D^\eta_{yy}$ and $D$, which start at O$(1)$, and the two coefficients 
$C_0, D_0$ are of O$(a^2)$.
Therefore, the total number of independent parameters is thirteen besides 
$m_{\rm critical}$, $2 B$ and $f$.

%
\subsection{O$(\tilde M a)$ term at NLO}
%
In eqs. (\ref{eq:final_pi}), (\ref{eq:final_K}) and  
(\ref{eq:final_eta}),
O$(\tilde M a)$ terms are considered at LO.
In this subsection  we present the results for the case that
these terms are treated as  NLO corrections.
First of all, the expressions for the shifted quark masses simplify to
\beqa
\tilde m &=& m - \frac{N c_2 + \tilde c_2}{2B} = m - m_{\rm  
critical}, \\
\tilde m_s &=& m_s - \frac{N c_2 + \tilde c_2}{2B}
= m_s - m_{\rm critical} .
\eeqa
The quark mass dependence of the pseudo scalar meson masses becomes
\beqa
m_{\pi, \rm total}^2 &=& (x+2y)[1-Nc_3-\tilde c_3]
-x N c_3
  +\frac{1}{f^2}\left[
  L_\pi^r \{ \frac{1}{2} x +  y + 5 C\}
+L_K^r\  4 C
+L_\eta^r \{ \frac{1}{2} x - \frac{1}{3} y +  C\}\right.\nn \\
&-&\left.
\{ (C_0+D_0)x + 2 C_0 y + (C_{\rm av}+D_{\rm av})x^2
+ D^\pi_{yy}y^2+2D xy\}
\right],
\label{eq:finalB_pi}\\
m_{K, \rm total}^2 &=& (x-y )[1-Nc_3-\tilde c_3]
-x N c_3
+\frac{1}{f^2}\left[
  L_K^r\ 6 C
+L_\pi^r \{ -\frac{1}{4} x + 3 C\}
+L_\eta^r \{ \frac{1}{3}(x - y) +  C\}\right.\nn\\
&-&\left.
\{ (C_0+D_0)x - C_0 y + (C_{\rm av}+D_{\rm av})x^2
+ D^K_{yy}y^2-D xy\}\right],
\label{eq:finalB_K}\\
m_{\eta, \rm total}^2 &=& (x-2y )[1-Nc_3-\tilde c_3]
-x N c_3
+\frac{1}{f^2}\left[
  L_\eta^r \{ -\frac{1}{2} x + \frac{5}{3} y + 3 C\}
+L_\pi^r \{ -\frac{1}{2} x - y + 3 C\}\right.\nn\\
&+&\left.
L_K^r \{ \frac{4}{3}(x - y) +  4 C\}
-
\{ (C_0+D_0)x - 2 C_0 y + (C_{\rm av}+D_{\rm av})x^2
+ D^\eta_{yy}y^2-2 D xy\}
\right] ,
\label{eq:finalB_eta}
\eeqa
where
\beqa
C&=&\frac{Z}{6}=\frac{ (9-N)c_2 + 3\tilde c_2}{6}, \quad
D^\pi_{yy}= 16 L_5+16 L_8 = 4 D^K_{yy},\nn\\
D^\eta_{yy}&=& 16 L_5+48 L_8 + 96 L_7
, \quad
D = C_{\rm av} +4 L_{5} + 8L_8, \nn \\
C_{\rm av} &=& 4(N L_6 + N L_4 + L_5),\quad
D_{\rm av} = 4 L_8 , \nn \\
C_0 &=& 4(W_0+W_1+W_2+W_3+W_4)+2W_5 +8N^2 W_6 +4NW_7  
+16NW_8+18W_9\nn\\
&-&8(W_{11}+W_{12})=a^2 W_C, \nn\\
D_0 &=& N[16(NW_6+W_7)+4W_8 +2W_{10}]=a^2 W_D . \nn
\eeqa
The number of independent parameters is reduced compared to the result given in the previous section. Besides $m_{\rm critical}$, $2 B$ and $f$
there are $c_3$, $\tilde c_3$, $C$, $L_4+L_6$, $L_5+L_8$
(note that $C_{\rm av}+D_{\rm av}$, $D^\pi_{yy}$, $D^K_{yy}$ and $D$ can be expressed by $L_4+L_6$ and $L_5+L_8$.)
and $D^\eta_{yy}$.
The total number of independent parameters besides $m_{\rm critical}$, $2 B$  
and $f$
is reduced  from 13 to 6. 

However, for improved theories there is some lattice spacing dependence implicit in the definition of the renormalized quark mass. This results in 3 parameters ($ b_B+ b_m^{(2)}$, $b_m^{(1)}$, $ b_m^{(3)}$),
as we will show in the next subsection, where we discuss O$(a)$ improved theories.
%
\subsection{Formula in O$(a)$ improved theories}
\label{Oaimprovedtheories}
%
We finally consider the case that a non-perturbatively O$(a)$ improved  
quark
action ({\it i.e.} the clover quark action ) is used in the lattice  
simulations \cite{Symanzik:1983dc,Symanzik:1983gh,Sheikholeslami:1985ij,Luscher:1996sc,Luscher:1996ug}.
In this case there are no on-shell O$(a)$ terms
in the Symanzik's effective theory provided that the relevant improvement coefficients are tuned non-perturbatively to an appropriate value. 
In particular,  O($a$) improvement requires that some $a$ dependence is absorbed in the definition of renormalized masses and the gauge coupling:
\beqa
m &\rightarrow&  m+b_m^{(1)} m^2 a + b_m^{(2)} (2m + m_s) m a + b_m^{(3)}  
(2m^2+m_s^2)a,\\
m_s &\rightarrow&  m_s+b_m^{(1)} m_s^2 a + b_m^{(2)} (2m + m_s) m_s a +
b_m^{(3)} (2m^2+m_s^2)a,\\
g_0^2 &\rightarrow&  g_0^2 \left( 1+b_g \frac{(2m+m_s) a}{3}\right),
\eeqa
where $b_g$ and $b_m = b_m^{(1)}+3(b_m^{(2)}+b_m^{(3)})$ are improvement  
coefficients
defined in Ref. \cite{Luscher:1996sc,Luscher:1996ug}.
Therefore, as long as on-shell quantities are considered,
there are no terms of O$(a)$, O$(\tilde M a)$, O$(\tilde M^2 a)$ etc.\ 
in the WChPT Lagrangian, if we replace
\beqa
\tilde m &\rightarrow& \bar m=\tilde m+ b_m^{(1)} \tilde m^2 a +  b_m^{(2)}  
(2\tilde m
+ \tilde m_s) \tilde m a +  b_m^{(3)} (2\tilde m^2+\tilde m_s^2)a,\\
\tilde m_s &\rightarrow& \bar m_s
=  \tilde m_s+ b_m^{(1)} \tilde m_s^2 a +  b_m^{(2)} (2\tilde m
+ \tilde m_s) \tilde m_s a +  b_m^{(3)} (2\tilde m^2+\tilde m_s^2)a, \\
B &\rightarrow& \bar B= B[ 1 +  b_B (2\tilde m + \tilde m_s)a ].
\eeqa
Here the last modification comes from the mass dependence of $g_0^2$
in the Symanzik's effective theory.

We emphasize that there are no terms linear in $a$ in the chiral Lagrangian and in the results for the pseudo scalar masses as long as one parameterizes it in terms of $\bar m$, which absorbs some O($a$) dependence through proper renormalization. Using $\tilde m$ instead, which is simpler in practice since this mass is directly proportional to the difference between the bare and the critical quark mass, there is some O($a$) dependence left explicit. 

Having made these remarks we can write down the WChPT expressions for non-perturbatively O($a$) improved theories:
\beqa
m_{\pi, \rm total}^2 &=& \bar x+2\bar y
  +\frac{1}{f^2}\left[
  L_\pi^r \{ \frac{1}{2} x +  y + 5 C\}
+L_K^r\  4 C
+L_\eta^r \{ \frac{1}{2} x - \frac{1}{3} y +  C\}\right.\nn \\
&-&\left.
\{ (C_0+D_0)x + 2 C_0 y + (C_{\rm av}+D_{\rm av})x^2
+ D^\pi_{yy}y^2+2D xy\}
\right],
\label{eq:finalC_pi}\\
m_{K, \rm total}^2 &=& \bar x-\bar y
+\frac{1}{f^2}\left[
  L_K^r\ 6 C
+L_\pi^r \{ -\frac{1}{4} x + 3 C\}
+L_\eta^r \{ \frac{1}{3}(x - y) +  C\}\right.\nn\\
&-&\left.
\{ (C_0+D_0)x - C_0 y + (C_{\rm av}+D_{\rm av})x^2
+ D^K_{yy}y^2-D xy\}\right],
\label{eq:finalC_K}\\
m_{\eta, \rm total}^2 &=& \bar x-2 \bar y
+\frac{1}{f^2}\left[
  L_\eta^r \{ -\frac{1}{2} x + \frac{5}{3} y + 3 C\}
+L_\pi^r \{ -\frac{1}{2} x - y + 3 C\}
+
L_K^r \{ \frac{4}{3}(x - y) +  4 C\} \right.\nn\\
&-&\left.
\{ (C_0+D_0)x - 2 C_0 y + (C_{\rm av}+D_{\rm av})x^2
+ D^\eta_{yy}y^2-2 D xy\}
\right] ,
\label{eq:finalC_eta}
\eeqa
where
\beqa
\bar x&=& \frac{2\bar B}{3}(2\bar m+\bar m_s)=
x[ 1 + ( b_B+ b_m^{(2)}) (2\tilde m+\tilde m_s)a]
+( b_m^{(1)}+3 b_m^{(3)})(2 \tilde m^2 +\tilde m_s^2)a, \\
\bar y &=& \frac{2\bar B}{6}(\bar m-\bar m_s)
= y[ 1+ ( b_B+ b_m^{(2)})(2\tilde m+\tilde m_s)a +  b_m^{(1)}( \tilde m+\tilde  
m_s)a],
\eeqa
and
\beqa
C&=&\frac{Z}{6}=\frac{ (9-N)c_2 + 3\tilde c_2}{6}, \quad
D^\pi_{yy}= 16 L_5+16 L_8 = 4 D^K_{yy},\nn\\
D^\eta_{yy}&=& 16 L_5+48 L_8 + 96 L_7
, \quad
D = C_{\rm av} +4 L_{5} + 8L_8, \nn \\
C_{\rm av} &=& 4(N L_6 + N L_4 + L_5),\quad
D_{\rm av} = 4 L_8 , \nn \\
C_0 &=& 4(W_0+W_1+W_2+W_3+W_4)+2W_5 +8N^2 W_6 +4NW_7  
+16NW_8+18W_9\nn\\
&-&8(W_{11}+W_{12})=a^2 W_C, \nn\\
D_0 &=& N[16(NW_6+W_7)+4W_8 +2W_{10}]=a^2 W_D . \nn
\eeqa
Independent parameters (besides $m_{\rm critical}$, $2 B$, and $f$)
are $ b_B+ b_m^{(2)}$, $ b_m^{(1)}$, $ b_m^{(3)}$, 
$C$, $L_4+L_6$, $L_5+L_8$ and $D^\eta_{yy}$.
The number of independent parameters besides $m_{\rm critical}$, $2 B$  
and $f$
is therefore 7, reduced from previously found 13.

%
\section{Concluding remarks}
%
In this paper we computed the pseudo scalar masses in 2+1 flavor WChPT.
We presented results for three different order countings, appropriate for various sizes of the O$(a\tilde M)$ term in the chiral Lagrangian. 
Depending on the lattice action used in the numerical simulation (unimproved, perturbatively improved, non-perturbatively improved) one has to choose one result for the chiral extrapolation. 
Since we have no prior knowledge about the size of the O$(a\tilde M)$ contribution we suggest to perform fits to the data with all three forms and let the data decide which form is most appropriate. 
 
The number of unknown fit parameters is significantly larger than in 2 flavor WChPT.
Using our results requires sufficiently enough data points in order to perform the chiral fits. The CP-PACS/JLQCD collaboration is currently performing 2+1 flavor simulations at three lattice spacings with five values for the light up and down quark mass and two different strange quark masses. At least for these simulations the number of data points exceeds the number of unknown fit parameters. Performing the chiral extrapolation of the CP-PACS/JLQCD data using our results is work in progress.     

%
%
\section*{Acknowledgments}
%
This work is supported in part by the Grants-in-Aid for
Scientific Research from the Ministry of Education,
Culture, Sports, Science and Technology
(Nos. 13135204, 15204015, 15540251, 16028201, 16$\cdot$11968
).
O.\ B.\ is supported in part by the University of Tsukuba Research Project.
S.\ T.\ is supported by Research Fellowships of
the Japan Society for the Promotion of Science for Young Scientists.
%
\appendix
%
\section{Useful formulae}
\label{sec:appA}
%

%
\subsection{Expansions in terms of $\pi$ }
%
In this subsection we collect some useful formulae necessary for the expansion of
the chiral Lagrangian in terms of the $\pi$ fields.
\subsubsection{LO terms}
At LO we have to expand terms at O$(\pi^4)$.
\beqa
&&\langle \partial_\mu \Sigma \partial_\mu\Sigma^\dagger\rangle
= \frac{4}{f^2}\langle \partial_\mu \pi \partial_\mu \pi \rangle
+\frac{8}{3f^4}\langle \partial_\mu \pi [\pi, \partial_\mu \pi] \pi  
\rangle,
\nn \\
&&\langle M\Sigma + \Sigma^\dagger M\rangle
=\langle 2 M\rangle -\frac{4}{f^2}\langle M\pi^2\rangle
+ \frac{4}{3f^4}\langle M\pi^4\rangle, \quad
\langle \Sigma + \Sigma^\dagger \rangle
=\langle 2\rangle -\frac{4}{f^2}\langle \pi^2\rangle
+ \frac{4}{3f^4}\langle \pi^4\rangle,\nn\\
&&\langle \Sigma + \Sigma^\dagger \rangle^2
= \langle 2 \rangle^2 -\frac{8}{f^2}\langle 2\rangle \langle  
\pi^2\rangle
+ \frac{8}{3f^4}\langle 2 \rangle\langle \pi^4\rangle
+\frac{16}{f^4}\langle \pi^2\rangle^2, \quad
\langle (\Sigma + \Sigma^\dagger)^2 \rangle
= \langle 4 \rangle -\frac{16}{f^2}\langle \pi^2\rangle
+ \frac{64}{3f^4}\langle \pi^4\rangle, \nn\\
&&\langle \Sigma +\Sigma^\dagger-2\rangle
\langle \partial_\mu \Sigma \partial_\mu\Sigma^\dagger\rangle
=-\frac{16}{f^4}\langle\pi^2\rangle \langle \partial_\mu \pi  
\partial_\mu \pi \rangle, \quad
\langle (\Sigma +\Sigma^\dagger-2)\partial_\mu \Sigma  
\partial_\mu\Sigma^\dagger\rangle
=-\frac{16}{f^4}\langle\pi^2 \partial_\mu \pi \partial_\mu \pi \rangle,
\nn \\
&&\langle \Sigma +\Sigma^\dagger\rangle \langle M\Sigma +  
\Sigma^\dagger M\rangle
=\langle 2\rangle \langle 2M\rangle-\frac{8}{f^2}\langle M\rangle  
\langle \pi^2\rangle
-\frac{8}{f^2}\langle 1\rangle \langle M \pi^2\rangle \nn\\
&&\hspace{5cm}+\,\frac{16}{f^4}\langle \pi^2\rangle \langle M\pi^2\rangle
+\frac{8}{3f^4}\langle M\rangle\langle\pi^4\rangle
+\frac{8}{3f^4}\langle 1\rangle \langle M\pi^4\rangle, \nn\\
&&\langle (\Sigma +\Sigma^\dagger)(M\Sigma + \Sigma^\dagger M)\rangle
=\langle 4M\rangle-\frac{16}{f^2}\langle M\pi^2\rangle
+\frac{64}{3f^4}\langle M\pi^4\rangle, \nn \\
&&\langle \Sigma - \Sigma^\dagger \rangle^2 = 0 , \quad
\langle (\Sigma - \Sigma^\dagger)^2 \rangle
= -\frac{16}{f^2} \langle \pi^2\rangle
+ \frac{64}{3f^4}\langle \pi^4\rangle,\nn\\
&&\langle \Sigma -\Sigma^\dagger\rangle \langle M\Sigma -  
\Sigma^\dagger M\rangle
=\frac{32}{3f^4}\langle \pi^3 \rangle \langle M \pi\rangle , \quad
\langle (\Sigma -\Sigma^\dagger)(M\Sigma - \Sigma^\dagger M)\rangle
=-\frac{16}{f^2}\langle M\pi^2\rangle
+\frac{64}{3f^4}\langle M\pi^4\rangle . \nn
\eeqa

\subsubsection{NLO terms}
We have to expand the NLO terms to O$(\pi^2)$.
\beqa
\langle \Sigma_{\mu\mu}\rangle
\langle M\Sigma +\Sigma^\dagger M\rangle &=& \frac{4}{f^2}\langle  
\partial_\mu\pi \partial_\mu\pi
\rangle \langle 2 M \rangle, \quad
\langle \Sigma_{\mu\mu} (M\Sigma +\Sigma^\dagger M)\rangle
= \frac{4}{f^2}\langle \partial_\mu\pi \partial_\mu\pi 2 M \rangle, \nn\\
\langle M\Sigma +\Sigma^\dagger M\rangle^2
&=& \langle 2M\rangle^2 -\frac{8}{f^2}\langle 2M\rangle\langle  
M\pi^2\rangle
, \quad
\langle M\Sigma -\Sigma^\dagger M\rangle^2
=-\frac{16}{f^2}\langle M\pi\rangle^2, \nn\\
\langle M\Sigma M\Sigma + \Sigma^\dagger M\Sigma^\dagger M\rangle
&=&-\frac{8}{f^2}\langle M\pi M\pi + M^2\pi^2 \rangle , \quad
\langle \Sigma^2 + (\Sigma^\dagger)^2 \rangle
=\langle 2\rangle -\frac{16}{f^2}\langle \pi^2  \rangle, \nn \\
\langle M \Sigma^2 + (\Sigma^\dagger)^2 M \rangle
&=&\langle 2M\rangle -\frac{16}{f^2}\langle M\pi^2  \rangle, \quad
\langle M \Sigma^3 + (\Sigma^\dagger)^3 M \rangle
=\langle 2M\rangle -\frac{36}{f^2}\langle M\pi^2  \rangle, \nn \\
\langle M^2 \Sigma + \Sigma^\dagger M^2 \rangle
&=&\langle 2M^2\rangle -\frac{4}{f^2}\langle M^2\pi^2  \rangle, \quad
\langle M\Sigma M\Sigma^2 + (\Sigma^\dagger)^2 M\Sigma^\dagger M\rangle
=\langle 2M^2\rangle -\frac{4}{f^2}\langle 4 M\pi M\pi + 5 M^2\pi^2 \rangle, \nn \\
\langle M \Sigma - \Sigma^\dagger M \rangle
&=&\frac{4i}{f}\langle M\pi\rangle , \quad
\langle M \Sigma^2 - (\Sigma^\dagger)^2 M \rangle
=\frac{8i}{f}\langle M\pi\rangle, \nn \\
\langle (\p_\mu\Sigma)^2 + (\p_\mu\Sigma^\dagger)^2 \rangle
&=&\langle (\p_\mu\Sigma)^2(\Sigma^\dagger)^2
+ \Sigma^2 (\p_\mu\Sigma^\dagger)^2 \rangle
=-\frac{8}{f^2}\langle \p_\mu\pi \p_\mu\pi \rangle. \nn
\eeqa

\subsection{Formula for traces}
After expanding the Lagrangian in terms of the $\pi$ fields, we have to take
the trace in the flavor space.
\subsubsection{LO terms}
\beqa
\langle 1 \rangle &=& N , \quad
\langle M_q \rangle = N M_0 , \quad
\langle \pi^2\rangle = \frac{1}{2}\sum_a \pi_a^2 , \quad\nn\\
2B \langle M_q \pi^2\rangle &=& \frac{1}{2}\sum_a m_a^2 \pi_a^2  , \quad
m_a^2 =\left\{
\begin{array}{ll}
m_\pi^2 = 2 B m, & a=1,2,3, \\
m_K^2 = B (m+m_s), & a=4,5,6,7, \\
m_\eta^2 = \dfrac{2B}{3}(m+2m_s), & a=8,\\
\end{array}
\right. \nn\\
2B \langle M_q \pi^4 \rangle &=&  
\sum_{a,b,c,d}F^{abcd}\pi_a\pi_b\pi_c\pi_d
, \quad
\langle \pi^4 \rangle = \frac{1}{4N}\sum_{a,b}\pi_a^2 \pi_b^2
+\frac{1}{8}\sum_{a\sim e}d^{abe}d^{cde}\pi_a\pi_b\pi_c\pi_d, \nn \\
  4 F^{abcd}&=& \frac{2BM_0}{2}\left\{\frac{2}{N}\delta^{ab}\delta^{cd}+
\sum_e d^{abe}d^{cde}\right\}
+  
\frac{2BM_8}{4}\left\{\frac{2}{N}(\delta^{ab}d^{cd8}+d^{ab8}\delta^{cd})
+\sum_{e} d^{ee8}d^{abe}d^{cde}\right\}, \nn \\
\langle \partial_\mu \pi \partial_\mu \pi\rangle &=&  
\frac{1}{2}\sum_a\partial_\mu\pi_a\partial_\mu\pi_a
, \quad
\langle \pi^2 \partial_\mu\pi \partial_\mu \pi \rangle =
\frac{1}{4N}\sum_{a,b}\pi_a^2 \partial_\mu \pi_b \partial_\mu \pi_b
+\frac{1}{8}\sum_{a\sim e}d^{abe}d^{cde}\pi_a\pi_b\partial_\mu  
\pi_c\partial_\mu \pi_d, \nn \\
\langle \partial_\mu \pi [\pi,\partial_\mu\pi] \pi \rangle &=&
-\frac{1}{4}\sum_{a\sim e} f^{abe}f^{cde} \partial_\mu\pi_a\pi_b
\partial_\mu\pi_c\pi_d
, \quad
\langle \pi^3 \rangle = \frac{1}{4}\sum_{a,b,c}d^{abc}\pi_a \pi_b \pi_c
, \quad
2B \langle M_q \pi \rangle = \frac{1}{2}2B M_8 \pi_8 .\nn
\eeqa

\subsubsection{NLO terms}
\beqa
\langle 2B\tilde M_q\rangle &=& N \tilde m_{\rm av}^2
, \quad
\langle \pi^2\rangle =\frac{1}{2}\sum_a \pi_a^2, \nn \\
\langle 2B\tilde M_q\rangle^2 &=& N^2 (\tilde m_{\rm av}^2)^2
, \quad
\langle (\partial_\mu\pi)^22B\tilde M_q\rangle =
\tilde m_{\rm av}^2 \frac{1}{2}\sum_a (\partial_\mu \pi_a)^2
+\Delta \tilde m^2 \frac{1}{4} \sum_a d^{aa8}(\partial_\mu\pi_a)^2,
\nn \\
\langle (2B\tilde M_q)^2\rangle &=& N (\tilde m_{\rm av}^2)^2+\frac{1}{2}
(\Delta\tilde m^2)^2\nn
, \quad
\langle 2B\tilde M_q\pi^2\rangle = \frac{1}{2}\sum_a \tilde m_a^2\pi_a^2
, \quad
\langle 2B\tilde M_q\pi\rangle^2 = \frac{1}{4} (\Delta \tilde m^2)^2\pi_8^2,\nn\\
2\langle (2B\tilde M_q\pi)^2\rangle &=& (\tilde m_{\rm av}^2)^2\sum_a\pi_a^2
+\tilde m_{\rm av}^2 \Delta\tilde m^2 \sum_a d^{aa8}\pi_a^2
+\frac{(\Delta\tilde m^2)^2}{4}\left[\frac{2}{N}\pi_8^2
+\sum_a\{(d^{aa8})^2-\sum_b(f^{ab8})^2\}
\pi_a^2\right], \nn\\
\langle (2B\tilde M_q)^2\pi^2\rangle &=& \frac{1}{2} (\tilde m_{\rm  
av}^2)^2\sum_a\pi_a^2
+\frac{1}{2} \tilde m_{\rm av}^2 \Delta\tilde m^2 \sum_a d^{aa8}\pi_a^2
+(\Delta\tilde m^2)^2\sum_a  
\left\{\frac{1}{4N}-\frac{1}{8\sqrt{3}}d^{aa8}\right\}\pi_a^2 .
\nn
\eeqa

\subsection{Group factors}
We have to calculate some Lie group factors.
\beqa
C^{ab}&\equiv& \sum_c f^{abc}f^{abc} = (1-\delta^{ab})C^{AB}=C^{ba},
  \nn \\
C^{\pi\pi}&=&C^{K_1K_1}=C^{K_2K_2}=1,\quad C^{\pi K}=C^{K_1K_2}
=1/4,\quad C^{\pi\eta}=0,\quad C^{K\eta}=3/4,\nn
\eeqa
where $\pi$ represents $a=1,2,3$, $K$ represents $a=4,5$ ($K_1$) and  
$a=6,7$
($K_2$) and $\eta$ represents $a=8$.
\beqa
D^{ab}&\equiv& \sum_c d^{aac}d^{bbc}=d^{aa3}d^{bb3}+d^{aa8}d^{bb8}
=D^{ba},\nn\\
D^{\pi\pi}&=&D^{K_1K_1}=D^{K_2K_2}=D^{\eta\eta}=1/3, \quad
D^{\pi K}=D^{K_1K_2}=-1/6, \quad
D^{\pi\eta}=-1/3,\quad D^{K\eta}=1/6 . \nn
\eeqa
\beqa
\tilde D^{ab}&\equiv& \sum_c d^{abc}d^{abc}=\tilde D^{ba}, \nn\\
\tilde D^{aa}&=&1/3, \quad
\tilde D^{\pi K}=\tilde D^{K_1 K_2}=1/4, \quad
\tilde D^{\pi \eta}=1/3, \quad
\tilde D^{K \eta}=1/12, \quad
\mbox{ others } = 0 . \nn
\eeqa
\beqa
E^{ab}&\equiv& \sum_c d^{cc8}d^{aac}d^{bbc}=
\frac{1}{\sqrt{3}}[d^{aa3}d^{bb3}-d^{aa8}d^{bb8}]=E^{ba},\nn\\
E^{\pi\pi}&=&E^{K_1K_2}=E^{\eta \eta}=-\frac{1}{3\sqrt{3}}, \quad
E^{\pi K}=E^{K_1K_1}=E^{K_2K_2}=\frac{1}{6\sqrt{3}}, \quad
E^{\pi \eta}=\frac{1}{3\sqrt{3}}, \quad
E^{K \eta}=-\frac{1}{6\sqrt{3}} . \nn
\eeqa
\beqa
\tilde E^{ab}&\equiv& \sum_c d^{cc8}d^{abc}d^{abc} =\tilde E^{ba},  
\qquad
\tilde E^{aa}
=\frac{1}{\sqrt{3}}\left\{
\begin{array}{cc}
-1/3 & \mbox{ for } \pi, \\
1/6  & \mbox{ for } K, \\
-1/3  & \mbox{ for } \eta,\\
\end{array}
\right. \nn \\
\tilde E^{\pi K}&=&-\frac{1}{8\sqrt{3}}, \quad
\tilde E^{\pi \eta}=\frac{1}{3\sqrt{3}}, \quad
\tilde E^{K_1 K_2}=\frac{1}{4\sqrt{3}}, \quad
\tilde E^{K \eta}=-\frac{1}{24\sqrt{3}}, \quad
\mbox{ others } = 0 . \nn
\eeqa
With these definition we have
\beqa
4F^{aabb}&=& 2B M_0\left\{\frac{1}{N}+\frac{1}{2}D^{ab}\right\}
          +  2B  
M_8\left\{\frac{1}{2N}(d^{aa8}+d^{bb8})+\frac{1}{4}E^{ab}\right\},\nn\\
4F^{abab}&=& 2B M_0\left\{\frac{1}{N}\delta^{ab}+\frac{1}{2}\tilde  
D^{ab}\right\}
          +  2B  
M_8\left\{\frac{1}{N}\delta^{ab}d^{aa8}+\frac{1}{4}\tilde  
E^{ab}\right\}.\nn
\eeqa

The following formulae for $N=3$ are useful.
\beqa
\frac{1}{N}+\frac{1}{2}D^{ab}+\tilde D^{ab}&=&
\left(
\begin{array}{cccccccc}
5/6 &1/2 & 1/2 & 1/2 & 1/2 & 1/2 & 1/2 & 1/2 \\
1/2 &1/2 & 1/2 & 5/6 & 1/2 & 1/2 & 1/2 & 1/2 \\
1/2 &1/2 & 1/2 & 1/2 & 1/2 & 1/2 & 1/2 & 5/6 \\
\end{array}
\right),\nn\\
\frac{1}{2N}(d^{aa8}+d^{bb8})+\frac{1}{4}E^{ab}+\frac{1}{2}\tilde  
E^{ab}&=&
\frac{1}{4\sqrt{3}}\left(
\begin{array}{cccccccc}
1/3 & 1   & 1   & 1/4 & 1/4 & 1/4 & 1/4 & 1   \\
1/4 & 1/4 & 1/4 &-1/6 &-1/2 &-1/2 &-1/2 &-5/4 \\
1   & 1   & 1   &-5/4 &-5/4 &-5/4 &-5/4 &-7/3 \\
\end{array}
\right),\nn
\eeqa
where $a=1,4,8$.

\subsection{1-loop contractions}
The following contraction formula is useful for the calculation of the
meson propagators at 1-loop.
\beqa
\langle \partial_\mu\pi [\pi,\partial_\mu\pi]\pi\rangle
&\rightarrow&
-\frac{1}{4}\sum_{a,b} C^{ab}
\left[ (\partial_\mu \pi_a)^2 I_0(\tilde m_b^2)
+\pi_a^2 I_1(\tilde m_b^2)\right], \quad
C^{ab}=\sum_c f^{abc}f^{abc}, \nn \\
I_0(m^2) &=&-\frac{m^2}{16\pi^2}\left[\Delta +1-\log m^2\right],
\quad \Delta =\frac{2}{\epsilon}-\gamma+\log(4\pi)
, \quad
I_1(m^2)= -m^2 I_0(m^2), \nn \\
\langle 2BM\pi^4\rangle &\rightarrow&
\sum_{a,b}\pi_a^2 2 I_0(\tilde m_b^2) [ F^{aabb}+2F^{abab} ],\nn \\
\langle \pi^4 \rangle &\rightarrow&
\sum_{a,b}\pi_a^2 I_0(\tilde m_b^2)\left[
\frac{1}{2N}(1+2\delta_{ab})+\frac{1}{4}\sum_e(d^{aae}d^{bbe}
+2 d^{abe}d^{abe})\right],\nn\\
\langle \pi^2\rangle^2 &\rightarrow&
\frac{1}{2}\sum_{a,b} \pi_a^2 I_0(\tilde m_b^2)(1+2\delta_{ab}),\nn \\
\langle \pi^2\rangle \langle 2BM \pi^2\rangle  &\rightarrow&
\frac{1}{4}\sum_{a,b} \pi_a^2 I_0(\tilde  
m_b^2)(m_a^2+m_b^2)(1+2\delta_{ab}),\nn \\
\langle \pi^2\rangle \langle (\partial_\mu\pi)^2\rangle  &\rightarrow&
\frac{1}{4}\sum_{a,b}\left[(\partial_\mu \pi_a)^2I_0(\tilde m_b^2)
+\pi_a^2 I_1(\tilde m_b^2)\right],\nn\\
\langle \pi^2 (\partial_\mu\pi)^2\rangle  &\rightarrow&
\sum_{a,b}\left[(\partial_\mu \pi_a)^2I_0(\tilde m_b^2)
+\pi_a^2 I_1(\tilde m_b^2)\right]
(\frac{1}{4N}+\frac{1}{8}\sum_e d^{aac}d^{bbc}),\nn\\
\langle \pi^3 \rangle \langle 2 BM\pi \rangle &\rightarrow&
\frac{3}{8} 2BM_8 d^{aa8}[ I_0(\tilde m_a^2)\pi_8^2 +I_0(\tilde  
m_8^2)\pi_a^2]. \nn
\eeqa

\section{Calculation of the pseudo scalar meson propagator at 1-loop}
\label{sec:appB}
In this appendix we give some details of the calculation of the pseudo scalar meson  
propagator at 1-loop in the WChPT.

\subsection{Effective action for BG fields}
In order to calculate meson masses at 1-loop, we use the  
background (BG) field method. We first split the $\pi$ field as
\beqa
\pi &=& \pi_Q + \pi_G,
\eeqa
where $\pi_Q$ represents quantum field while $\pi_G$ is a BG field which
satisfies the equation of motion. Inserting this into the chiral effective  
Lagrangian, we have
\beqa
L_{\rm LO}(\pi)&=& L_{\rm LO}(\pi_G) + L^{(2)}(\pi_Q) +  
L^{(4)}(\pi_Q+\pi_G)-L^{(4)}(\pi_G)  .
\eeqa
Integrating out $\pi_Q$ we obtain the following formula
\beqa
e^{-S_{\rm eff}(\pi_G)}&=& \int {\cal D}\pi_Q e^{-\int d^4x L_{\rm  
LO}(\pi)}\nn\\
&=& e^{-\int d^4 x  L_{\rm LO}(\pi_G)} \int  {\cal D}\pi_Q e^{-\int  
d^4x  L^{(2)}(\pi_Q)}
e^{-\int d^4x (L^{(4)}(\pi_Q+\pi_G)-L^{(4)}(\pi_G))}.
\eeqa
This leads to
\beqa
S_{\rm eff}(\pi_G)&=& \int d^4x L_{\rm LO}(\pi_G) - \log  \langle
e^{-\int d^4x (L^{(4)}(\pi_Q+\pi_G)-L^{(4)}(\pi_G))} \rangle,
\eeqa
where
\beqa
\langle f(\pi_Q,\pi_G) \rangle &=& \int  {\cal D}\pi_Q e^{-\int d^4x   
L^{(2)}(\pi_Q)} f(\pi_Q,\pi_G) .
\eeqa
Expanding $S_{\rm eff}$ in terms of $\pi_G$, we obtain
\beqa
S_{\rm eff}(\pi_G)&=&\mbox{ constant } + S_{\rm eff}^{(2)}(\pi_G)  
+\sum_{n=3}^\infty S_{\rm eff}^{(n)}(\pi_G),
\eeqa
where $S_{\rm eff}^{(n)}(\pi_G)$ contains the n-th power of the field  
$\pi_G$. In the calculation of the pseudo scalar masses we are
interested in the $n=2$ case. We write
\beqa
S_{\rm eff}^{(2)}(\pi_G) &=& \int d^4 x L_{\rm LO}(\pi_G)
+S_{\rm 1-loop}^{(2)}(\pi_G) + \cdots,
\eeqa
where $\cdots$ represent the higher loop contributions. We call $S_{\rm  
1-loop}^{(2)}$
the 1-loop contribution to the meson propagator and write
\beqa
S_{\rm 1-loop}^{(2)}(\pi_G)&=&\int d^4 x L_{\rm 1-loop}^{(2)}(\pi_G).
\eeqa

\subsection{Expansion of the LO Lagrangian}
We now expand the LO Lagrangian (\ref{eq:LO}) in terms of the
pseudo scalar  field $\pi_a$. Using the expansion and trace formulae given
in appendix \ref{sec:appA},
we obtain
\beqa
L^{(2)}&=&\langle (\partial_\mu \pi)^2\rangle+2B\langle M_q\pi^2\rangle
-\frac{c_2}{2}\langle 2\rangle\langle \pi^2\rangle
-\tilde c_2 \langle \pi^2\rangle \nn\\
&-& c_3\langle 2B M_q\rangle \langle \pi^2\rangle
- c_3\langle 1\rangle 2B \langle M_q \pi^2\rangle
- \tilde c_3 2B \langle M_q \pi^2\rangle \nn \\
&=&\frac{1}{2}\sum_a\left[ (\partial_\mu\pi_a)^2
+\tilde m_a^2 \pi_a^2 \right],
\label{eq:LO_2ptApp}
\eeqa
at second order in $\pi_a$,
where the pseudo scalar meson masses at  LO are given by
\beqa
\tilde m_a^2 &=& m_a^2(1-N c_3-\tilde c_3)-m_{\rm av}^2 N c_3
-Nc_2-\tilde c_2\,,\\
m_{\rm av}^2&=&\frac{1}{N^2-1}\sum_a m_a^2.
\eeqa
For $N=3$ flavors we have
\beqa
m_a^2 &=&\left\{
\begin{array}{ll}
m_\pi^2 = 2 B m, & a=1,2,3, \\
m_K^2 = B (m+m_s), & a=4,5,6,7, \\
m_\eta^2 = \dfrac{2B}{3}(m+2m_s), & a=8.\\
\end{array}
\right. \quad
m_{\rm av}^2 = 2B\frac{ 2m + m_s}{3},
\eeqa
Eq.(\ref{eq:LO_2ptApp}) gives the pseudo scalar meson propagator at LO.

The 4-th order terms in the LO Lagrangian become
\beqa
L^{(4)}&=&\frac{2}{3f^2}\langle \partial_\mu  
\pi[\pi,\partial_\mu\pi]\pi\rangle
-\frac{1-Nc_3-4\tilde c_3}{3f^2}2B\langle M\pi^4\rangle \nn\\
&+&\frac{Nc_2+4\tilde c_2+Nc_3 m_{\rm av}^2}{3f^2}\langle  
\pi^4\rangle
+\frac{c_2}{f^2}\langle \pi^2\rangle^2+\frac{2c_3}{f^2}\langle  
\pi^2\rangle
\langle 2BM\pi^2\rangle \nn\\
&-&\frac{4c_0}{f^2}\langle \pi^2\rangle \langle (\partial_\mu  
\pi)^2\rangle
-\frac{4\tilde c_0}{f^2}\langle \pi^2 (\partial_\mu \pi)^2\rangle
+\frac{4c_5}{3f^2}\langle \pi^3\rangle  \langle 2 B M\pi\rangle .
\label{eq:LO_4pt}
\eeqa
There terms give the 4-point interaction vertices of the pseudo scalar mesons.

\subsection{1-loop contribution to the propagator}
Using the formulae in appendix \ref{sec:appA},
it is now easy to calculate $ L_{\rm 1-loop}^{(2)}$ .
Including the tree level contribution we obtain
\beqa
S_{\rm eff}^{(2)}(\pi) &=& \int d^4 x\frac{1}{2}\sum_a Z_a\left[
(\partial_\mu\pi_a)^2 + m_{a,R}^2\pi_a^2 
\right]\,.
\eeqa
For the $\pi$ ($a=1,2,3$), $K$ ($a=4,5,6,7$) and $\eta$ ($a=8$) 
we find the wave function renormalization as
\beqa
Z_\pi &=& 1-\frac{1}{3f^2}\left[ L_\pi\{ 2+9(2c_0+\tilde c_0)\}
+L_K\{1+6(4c_0+\tilde c_0)\}+L_\eta (6c_0+\tilde c_0)\right],\nn\\
Z_K &=& 1-\frac{1}{3f^2}\left[ L_\pi\{ 3/4+9(2c_0+\tilde c_0/2)\}
+L_K\{3/2+3(8c_0+3\tilde c_0)\}+L_\eta (3/4+6c_0+5\tilde  
c_0/2)\right],\nn\\
Z_\eta &=& 1-\frac{1}{3f^2}\left[ L_\pi 3( 6c_0+\tilde c_0)
+L_K\{3+2(12c_0+5\tilde c_0)\}+L_\eta 3(2c_0+\tilde c_0)\right],
\eeqa
where
\beqa
L_\pi &=& I_0(\tilde m_\pi^2), \quad L_K = I_0(\tilde m_K^2), \quad
L_\eta = I_0(\tilde m_\eta^2) .
\eeqa
Similarly we have
\beqa
m_{\pi,R}^2 = \tilde m_\pi^2 &+&\frac{1}{3f^2}\left[
L_\pi\left\{C^\pi_{\pi} 2 \tilde m_\pi^2 -\frac{5}{2} X \tilde  
m_{\rm av}^2
-5Y \frac{\Delta\tilde m^2}{4\sqrt{3}} +\frac{5}{2} Z\right\}  
\right.\nn\\
&+&L_K\left\{C^\pi_{K} (\tilde m_K^2+ \tilde m_\pi^2) -2 X \tilde  
m_{\rm av}^2
-Y\frac{\Delta\tilde m^2}{4\sqrt{3}} +2 Z\right\} \nn\\
&+&\left.
L_\eta\left\{C^\pi_{\eta} (\tilde m_\eta^2+ \tilde m_\pi^2)  
-\frac{1}{2} X \tilde m_{\rm av}^2
-(Y-12c_5) \frac{\Delta\tilde m^2}{4\sqrt{3}} +\frac{1}{2} Z\right\}
\right], \label{mpidiv}\\
m_{K,R}^2 = \tilde m_K^2 &+&\frac{1}{3f^2}\left[
L_K\left\{C^K_{K} 2 \tilde m_K^2 -3 X \tilde m_{\rm av}^2
+3Y \frac{\Delta\tilde m^2}{4\sqrt{3}}+3 Z\right\} \right.\nn\\
&+&L_\pi\left\{C^K_{\pi} (\tilde m_\pi^2+ \tilde m_K^2) -\frac{3}{2}  
X \tilde m_{\rm av}^2
-\frac{3}{4}Y \frac{\Delta\tilde m^2}{4\sqrt{3}} +\frac{3}{2} Z\right\}  
\nn\\
&+&\left.
L_\eta\left\{C^K_{\eta} (\tilde m_\eta^2+ \tilde m_K^2) -\frac{1}{2}  
X \tilde m_{\rm av}^2
+\frac{5}{4}(Y-\frac{24}{5}c_5) \frac{\Delta\tilde m^2}{4\sqrt{3}}  
+\frac{1}{2} Z\right\}
\right], \\
m_{\eta,R}^2 = \tilde m_\eta^2 &+&\frac{1}{3f^2}\left[
L_\eta\left\{C^\eta_{\eta} 2 \tilde m_\eta^2 -\frac{3}{2} X \tilde  
m_{\rm av}^2
+5(Y-\frac{24}{5}c_5)\frac{\Delta\tilde m^2}{4\sqrt{3}}+\frac{3}{2}  
Z\right\} \right.\nn\\
&+&L_\pi\left\{C^\eta_{\pi} (\tilde m_\pi^2+ \tilde m_\eta^2)  
-\frac{3}{2} X \tilde m_{\rm av}^2
-3(Y-12c_5)\frac{\Delta\tilde m^2}{4\sqrt{3}} +\frac{3}{2} Z\right\}  
\nn\\
&+&\left.
L_K\left\{C^\eta_{K} (\tilde m_K^2+ \tilde m_\eta^2) -2 X \tilde  
m_{\rm av}^2
+5(Y-\frac{24}{5}c_5)\frac{\Delta\tilde m^2}{4\sqrt{3}} +2 Z\right\}
\right].\label{etadiv}
\eeqa
for the pseudo scalar meson mass. The parameters in these expressions have already been given in subsection \ref{subsec:quarkmassdependenceofmesonmasses}.

\subsection{NLO contribution to meson propagators}
Using the formulae for the expansion in powers of the pion field and the trace formulae in the appendix  
\ref{sec:appA},
we have
\beqa
L_{\rm NLO} &=& \frac{1}{2}Z_a^{\rm NLO}\left[(\p_\mu\pi_a )^2 +  
m_{a,\rm NLO}^2 \pi_a^2\right],
\eeqa
where the wave function renormalization factor is given by
\beqa
Z_a^{\rm NLO} &=& \frac{1}{f^2}\left[ \tilde m_{\rm av}^2 z_{\rm av} +  
\Delta \tilde m^2 z_\Delta^a
+ z_0\right], \nn\\
z_{\rm av} & =& 4 (N \tilde L_4 +\tilde L_5),\nn\\
z_\Delta^a& = & 2 d^{aa8}\tilde L_5,\nn\\
z_0 & =&  4(W_0+W_1+W_2+W_3+W_4) - 8(W_{11}+W_{12}), \nn
\eeqa
and the mass term is defined as
\beqa
m_{a,\rm NLO}^2 &=& -\frac{1}{f^2}\left[ \tilde m_a^2 \left\{  \tilde  
m_{\rm av}^2 C_{\rm av}
+\Delta \tilde m^2 C_\Delta^a + C_0 \right\}
+\tilde m_{\rm av}^2 \left\{ \tilde m_{\rm av}^2 D_{\rm av}+\Delta  
\tilde m^2 D_\Delta^a + D_0 \right\}
+(\Delta \tilde m^2)^2 E_\Delta^a\right],
\label{eq:mass_NLO}  \\
C_{\rm av} &=& 4(N\tilde L_6 + N\tilde L_4 +\tilde L_5),\quad
D_{\rm av} = 2(\tilde L_8 + \tilde L_8^\prime) + V_{\rm av},\quad
C_\Delta^a = 2 d^{aa8} \tilde L_5, \nn \\
D_\Delta^a &=& 2 d^{aa8}( \tilde L_8+\tilde L_8^\prime ), \quad
E_\Delta^a = e_a \tilde L_8 + e_a^\prime \tilde L_8^\prime + V_{\Delta} +\delta_{a8}
\left(2\tilde L_7 +\frac{1}{N}\tilde L_8^\prime \right),  \nn \\
C_0 &=& 4(W_0+W_1+W_2+W_3+W_4)+2W_5 +8N^2 W_6 +4NW_7  
+16NW_8+18W_9\nn\\
& &-\,8(W_{11}+W_{12})=a^2 W_C, \nn\\
D_0 &=& N[ 16(NW_6+W_7)+4W_8 +2W_{10}]=a^2 W_D . \nn
\eeqa
The constants here are given by
\beqa
\tilde L_4 &=& L_4 + V_0+V_1+V_2+V_3 = L_4 +a L_4^1,\\
\tilde L_5 &=& L_5 + V_4+V_5+V_6-V_{16}-V_{17} = L_5 +a L_5^1,\\
\tilde L_6 &=& L_6 + 2NV_8+\frac{1}{4}V_{10}+\frac{5}{2}V_{11} = L_6 +a  
L_6^1,
\\
\tilde L_7 & =& L_7 + 2NV_{14}+2V_{15} = L_7 +a L_7^1,\\
\tilde L_8 &=& L_8 + 2N V_9+\frac{1}{2}V_{12}+\frac{5}{2}V_{13}
= L_8 +a L_8^{1a}, \\
\tilde L_8^\prime &=& L_8 + 2N V_9+2V_{13}= L_8 +a L_8^{1b},
\eea
and
\bea
V_{\rm av} &=& N(V_7 + 2 V_9 + 4 N V_8), \quad
V_{\Delta} = \frac{1}{2}(V_7 + 2 V_9),\\
  e_{a}
  &=&
  \frac{1}{N} - \frac{1}{2\sqrt{3}} d^{aa8},
\mbox{\hspace{5mm}}
  (e_{\pi},e_{K},e_{\eta})
  =
  (\frac{1}{6},\frac{5}{12},\frac{1}{2}),
\\
  e_{a}^{\prime}
  &=&
  \frac{1}{2} [ (d^{aa8})^2 - \sum_b (f^{ab8})^2 ],
\mbox{\hspace{5mm}}
  (e_{\pi}^{\prime},e_{K}^{\prime},e_{\eta}^{\prime})
  =
  (\frac{1}{6},-\frac{1}{3},\frac{1}{6}).
\eeqa

\subsection{Cancellation of UV divergence}
In order to perform the renormalization,
we consider divergent parts of meson masses in eq.\ (\ref{mpidiv}) - \pref{etadiv},
which are given by
\beqa
[m_a^2 ]_{\rm div.}&=&
-\frac{\Delta}{48\pi^2f^2}\left[ C^a_x x+C^a_y y +C^a_{xx} x^2  
+C^a_{yy}y^2
+C^a_{xy}xy\right],
\eeqa
where $x= \tilde m_{\rm av}^2$ and $y=\frac{1}{4\sqrt{3}}\Delta\tilde  
m^2$,
in terms of which, $\tilde m_\pi^2=x+2y$, $\tilde m_K^2 = x-y$, and
$\tilde m_\eta^2 = x-2y$. Constants are given by
\beqa
C^\pi_x &=& C^K_x=C^\eta_x=5Z, \quad
C^\pi_y = 2 Z,\quad C^K_y=-Z, \quad C^\eta_y=-2Z ,\nn\\
C^a_{xx}&=& 2\sum_b C^a_{ba}-5X = 2(3+48c_0+16\tilde c_0 +30 c_3 B)-  
5X,\nn\\
C^\pi_{yy}&=& 8C^\pi_{\pi}-C^\pi_{K}-7Y-24c_5\tilde B =
  15 + 120 c_0 + 66 \tilde{c}_0
+ \tilde B [ - 7 + 129 c_3 + 28 \tilde{c}_{3} - 24 c_5 ], \nn\\
C^K_{yy}&=& 2C^K_{K}+2C^K_{\pi}+6C^K_{\eta}-7Y+12c_5\tilde B =
  9 + 120 c_0 + 42 \tilde{c}_0
+ \tilde B [ - 7 + 93 c_3 + 28 \tilde{c}_{3} + 12 c_5 ], \nn\\
C^\eta_{yy}&=& 8C^\eta_{\eta}+3C^\eta_{K}-21Y+144c_5\tilde B =
  9 + 120 c_0 + 54 \tilde{c}_0 
+ \tilde B [ - 21 + 171 c_3 + 84 \tilde{c}_{3} + 144 c_5 ], \nn\\
C^\pi_{xy}&=& 8C^\pi_{\pi}-C^\pi_{K}-4C^\pi_{\eta}-2X-7Y+12c_5 \tilde B =
  15 + 96 c_0 + 62 \tilde{c}_0 + 96 c_3 \tilde B 
 -2X - 7 Y + 12 c_5 \tilde B, \nn\\
C^K_{xy}&=& -4C^K_{K}+5C^K_{\pi}-7C^K_{\eta}+X+\frac{7}{2}Y-6c_5\tilde B =
-\frac{1}{2}C^\pi_{xy},\nn\\
C^\eta_{xy}&=& -8C^\eta_{\eta}+4C^\eta_{\pi}-5C^\eta_{K}
+2X+7Y-12c_5\tilde B = - C^\pi_{xy}.\nn
\eeqa
On the other hand, the NLO contributions 
 lead to
\beqa
[m_a^2 ]_{\rm NLO}&=&
-\frac{1}{f^2}\left[ D^a_x x+D^a_y y +D^a_{xx} x^2 +D^a_{yy}y^2
+D^a_{xy}xy\right] ,\nn
\eeqa
where
\beqa
D^a_x &=& C_0 + D_0, \quad D^\pi_y=2C_0, \quad D^K_y=-C_0, \quad
D^\eta_y=-2C_0, \quad
D^a_{xx}= C_{\rm av}+D_{\rm av},\nn\\
D^\pi_{yy}&=&16\tilde L_5 + 8(\tilde L_8 + \tilde L_8^\prime) + 48 V_{\Delta},\quad
D^K_{yy}=4\tilde L_5 + 20\tilde L_8 - 16 \tilde L_8^\prime + 48 V_{\Delta},\nn\\
D^\eta_{yy}&=&16\tilde L_5 + 24\tilde L_8+ 24\tilde L_8^\prime
+96\tilde L_7 + 48 V_{\Delta},\nn\\
D^\pi_{xy}&=& 2 C_{\rm av}+8\tilde L_5+8(\tilde L_8 +\tilde L_8^\prime)
=-2 D^K_{xy}= - D^\eta_{xy}.\nn
\eeqa
In order to cancel the UV divergences, the divergent part in the  NLO terms must be chosen according to
\beqa
\left[C_0\right]_{\rm div.} &=& -\frac{\Delta}{48\pi^2}Z, \quad
\left[D_0 \right]_{\rm div.} = -\frac{\Delta}{48\pi^2}4Z, \nn\\
\left[C_{\rm av}+ D_{\rm av}\right]_{\rm div.}
  &=& -\frac{\Delta}{48\pi^2}[ 2(3+48c_0+16\tilde c_0 +30 c_3 \tilde B)-5X],\nn \\
\left[16\tilde{L}_5 + 8 ( \tilde{L}_8 + \tilde{L}_8^{\prime}) + 48 V_{\Delta}\right]_{\rm div}
  &=&
 -\frac{\tilde{\Delta}}{48 \pi^2}
  [15 + 120 c_0 + 66 \tilde{c}_0
+ \tilde B ( - 7 + 129 c_3 + 28 \tilde{c}_{3} - 24 c_5 )],
\nn\\
\left[ 4\tilde{L}_5 + 20 \tilde{L}_8 -16 \tilde{L}_8^{\prime} + 48 V_{\Delta}\right]_{\rm div}
  &=&
 -\frac{\tilde{\Delta}}{48 \pi^2} 
  [9 + 120 c_0 + 42 \tilde{c}_0
+ \tilde B ( - 7 + 93 c_3 + 28 \tilde{c}_{3} + 12 c_5 )],
\nn\\
\left[ 16\tilde{L}_5 + 96\tilde{L}_7 + 24( \tilde{L}_8 + \tilde{L}_8^{\prime}) + 48 V_{\Delta}\right]_{\rm div}
  &=&
 -\frac{\tilde{\Delta}}{48 \pi^2} 
  [9 + 120 c_0 + 54 \tilde{c}_0 
+ \tilde B ( - 21 + 171 c_3 + 84 \tilde{c}_{3} + 144 c_5 )],\nn\\
\left[
2C_{\rm av}+8\tilde L_5+8\tilde L_8+8\tilde L_8^\prime \right]_{\rm  
div.}&=&
-\frac{\Delta}{48\pi^2}[15+96c_0+62\tilde c_0 + 96c_3 \tilde B-2X-7Y+12c_5 \tilde B].\nn
\eeqa
Notice that we can remove all divergences $[m_a^2]_{\rm div.}$
consistently by these parameters, as it should be.

%

\end{document}